\documentclass[12pt,a4paper]{article}
\usepackage[german, english]{babel}
\usepackage{a4wide}
\usepackage{amsmath}
\usepackage{amssymb}
\usepackage{ifthen}
\usepackage{epsfig}
\begin{document}
\newcommand{\beq}{\begin{equation}}
\newcommand{\eeq}{\end{equation}}
\newcommand{\lsim}{\mbox{$<$\hspace{-0.8em}\raisebox{-0.4em}{$\sim$}}}

\newcommand{\B}{\mbox{${\bf B}$}}
\newcommand{\E}{\mbox{${\bf E}$}}
\newcommand{\n}{\mbox{${\bf n}$}}
\newcommand{\p}{\mbox{${\bf p}$}}
\newcommand{\s}{\mbox{${\bf s}$}}

\newcommand{\bb}{\mbox{${\bf b}$}}
\newcommand{\bd}{\mbox{${\bf d}$}}
\newcommand{\bI}{\mbox{${\bf I}$}}
\newcommand{\bj}{\mbox{${\bf j}$}}
\newcommand{\bk}{\mathbf{k}}
\newcommand{\bq}{\mathbf{q}}
\newcommand{\br}{\mbox{${\bf r}$}}
\newcommand{\bv}{\mbox{${\bf v}$}}

\newcommand{\vA}{\mathbf{A}}
\newcommand{\va}{\mathbf{a}}
\newcommand{\vom}{\mbox{{\boldmath$\omega$}}}
\newcommand{\Na}{\mbox{{\boldmath$\nabla$}}}
\newcommand{\si}{\mbox{{\boldmath$\sigma$}}}

\newcommand{\al}{\mbox{${\alpha}$}}
\newcommand{\be}{\mbox{${\beta}$}}
\newcommand{\ga}{\mbox{${\gamma}$}}
\newcommand{\Ga}{\mbox{${\Gamma}$}}
\newcommand{\De}{\mbox{${\Delta}$}}
\newcommand{\la}{\mbox{${\lambda}$}}
\newcommand{\ep}{\mbox{${\varepsilon}$}}
\begin{titlepage}
\title{P and T odd nuclear moments}
\author{V.F. Dmitriev\thanks{E-mail: dmitriev@inp.nsk.su} \\ and \\
I.B. Khriplovich\thanks{E-mail: khriplovich@inp.nsk.su}}
\maketitle
\begin{abstract}
We discuss the nuclear anapole moment,
a new P odd, T even electromagnetic multipole. Its discovery in the cesium
experiment gives a first-rate information on parity-nonconserving nuclear
forces. New prospects in the field are considered. Upper limits on
the electric dipole moments, which are P odd, T odd multipoles, of
elementary particles and atoms are presented, and their physical
implications discussed. The atomic experiments are demonstrated to
be as informative in this respect as the neutron ones. Tremendous
progress in the field can be expected from experiments at ion
storage rings and linear electrostatic traps.
\end{abstract}
\end{titlepage}

\section{Introduction}
We discuss in the present article two sorts of electromagnetic
nuclear moments. To the first type belong those which violate the
invariance under P, the parity transformation, but conserve T,
time reversal invariance. To the second type belong the moments
violating both P and T.

P odd, T even moments arise in a system without centre of
inversion. Though their existence was predicted in the middle of
the past century \cite{zel}, the lowest of them, the so-called
anapole, that of the $^{133}$Cs nucleus, was experimentally
discovered only few years ago \cite{woo}. Being of a great
interest by itself, this discovery gives also quite interesting
and nontrivial information on P odd nuclear forces.

As to P odd, T odd multipoles, the lowest of them, electric dipole
moments (EDM), have interested physicists since 1950, when it was
first suggested that there was no experimental evidence that
nuclear forces are symmetric under parity
transformation~\cite{pura}.

In 1964 it was discovered that the invariance under CP
transformation, which combines charge conjugation with parity, is
violated in $K$-meson decays. This provided a new incentive for
EDM searches. Since the combined operations of CPT are expected to
leave a system invariant, breakdown of CP invariance should be
accompanied by T violation. Thus there is a reason to expect that
EDMs should exist at some level.

The original neutron EDM experiments were later supplemented with
searches for EDMs of other objects, nuclei included. These
investigations are pursued now by many groups. Over the years, the
upper limit on the neutron EDM has been improved by seven orders
of magnitude, and the upper limit on the electron EDM obtained in
atomic experiments is even more strict.

Even without the discovery of the effect sought, the neutron and
atomic experiments have ruled out most models of CP violation. As
to the mechanism of CP violation incorporated in the standard
model of electroweak interactions, which is most popular at
present, the predictions for the neutron and nuclear EDMs are
roughly six orders of magnitude below the present experimental
bound. The gap for the electron EDM is much larger.

But does this mean that the EDM experiments are of no serious
interest for the elementary particle physics, that they are
nothing but mere exercises in precision spectroscopy? Just the
opposite. It means that {\it the EDM experiments now, at the
present level of accuracy, are extremely sensitive to possible new
physics beyond the standard model, physics to which the kaon
decays are insensitive.}

\section{Nuclear anapole moments}

\subsection{General discussion. Anapole moment of $^{133}$Cs}

In a system which has no definite parity, a special distribution
of magnetic field may arise \cite{zel}. It cannot be reduced to
common electromagnetic multipoles, such as dipole or quadrupole
moments, but looks like the magnetic field created by a current in
toroidal winding. This special source of electromagnetic field was
called (by A.S. Kompaneets) anapole.

For many years the anapole remained a theoretical curiosity only.
The situation has changed with the investigations of parity
nonconservation (PNC) in atoms. The tiny P odd effects are
enhanced therein in particular by the weak nuclear charge $Q_W$,
which is close numerically to the neutron number and is thus about
a hundred in heavy atoms. However, this enhancement refers only to
the nuclear-spin-independent weak interaction. Meanwhile, the
nuclear-spin-dependent effects due to neutral currents not only
lack the mentioned coherent enhancement, but are also strongly
suppressed numerically in the electroweak theory. Therefore, the
observation of PNC nuclear-spin-dependent effects in atoms looked
absolutely unrealistic.

However, it was demonstrated \cite{fk,fks} that these effects in
heavy atoms are dominated not by the weak interaction of neutral
currents, but by the electromagnetic interaction of atomic
electrons with nuclear anapole moment (AM). It should be mentioned
first of all that the magnetic field of an anapole is contained
within it, in the same way as the magnetic field of a toroidal
winding is completely confined inside the winding. It means that
the electromagnetic interaction of an electron with the nuclear AM
occurs only as long as the electron wave function penetrates the
nucleus. In other words, this electromagnetic interaction is as
local as the weak interaction, they are quite similar. The nuclear
AM is induced by PNC nuclear forces and is therefore proportional
to the same Fermi constant $G=1.027\times 10^{-5} m^{-2}$ (we use
the units $\hbar=1, c=1$; $m$ is the proton mass), which
determines the magnitude of the weak interactions in general and
that of neutral currents in particular. The electron interaction
with the AM, being of the electromagnetic nature, introduces an
extra small factor into the effect discussed, the fine-structure
constant $\alpha=1/137$. Then, how it comes that this effect is
dominating? The answer follows from the same picture of a toroidal
winding. It is only natural that the interaction discussed is
proportional to the magnetic flux through such a winding, and
hence in our case to the cross-section of the nucleus, i.e. to
$A^{2/3}$, where $A$ is the atomic number. In heavy nuclei this
enhancement factor is close to 30 and compensates essentially for
the smallness of the fine-structure constant $\alpha$. As a
result, the dimensionless effective constant $\kappa$ which
characterizes the anapole interaction in the units of $G$ is not
so small in heavy atoms, but is numerically close to 0.3 (we use
the same definition of the effective constant $\kappa$ as
in~\cite{fk,fks}).

Still, the interaction discussed constitutes only about one
percent of the main atomic PNC effect, which is due to $Q_W$. To
single out the anapole interaction one should compare the PNC
effects for different hyperfine components of an optical
transition. The main effect, independent of the nuclear spin, will
obviously be the same for all components. But the anapole
contribution depends on the mutual orientation of the nuclear spin
and the electron total angular momentum, and thus changes from one
hyperfine component to another. The observation of this tiny
effect is an extremely difficult problem, and it is no accident
that the searches for the nuclear AM demanded many years of hard
work by several groups.

The nuclear anapole moment was experimentally discovered in 1997
\cite{woo}. This result for the total effective constant of the
PNC nuclear-spin-dependent interaction in $^{137}$Cs is
\beq
\kappa_{tot}=0.44(6).
\eeq
To extract this number from experimental data, the results of
atomic calculations \cite{fra,kra} were used; these calculations
were performed in different approaches, but are in excellent
agreement. Their accuracy is no worse than 2~--~3\%. If one
excludes the neutral current nuclear-spin-dependent contribution
from the above number, as well as the result of the combined
action of $Q_W$ and the usual hyperfine interaction, the answer
for the anapole constant will be
\beq\label{exp}
\kappa=0.37(6).
\eeq
Thus, the existence of an AM of the $^{137}$Cs nucleus is reliably
established. A beautiful new physical phenomenon, a peculiar
electromagnetic multipole has been discovered.

\subsection{Theoretical issues. Implications for P odd nuclear forces}

But the discussed result does not reduce to only this. It brings
valuable information on PNC nuclear forces. Of course, to this end
it should be combined with reliable nuclear calculations. However,
it is instructive to start, as it was done in \cite{fks}), with a
rather crude approximation. Not only one assumes here that the
nuclear spin $\mathbf{I}$ coincides with the total angular
momentum of an odd valence nucleon, while the other nucleons form
a core with the zero angular momentum. The next assumption is that
the core density $\rho(r)$ is constant throughout the space and
coincides  with the mean nuclear density $\rho_0$. The last
assumption is reasonable if the wave function of the external
nucleon is mainly localized in the region of the core. Then simple
calculations give the following result for the anapole
constant~\cite{fks}:
\begin{equation}\label{an}
\kappa=\frac{9}{10}\,g\,\frac{\alpha \mu}{m r_0}\, A^{2/3}.
\end{equation}
Here $g$ is the effective constant of the P odd interaction of the
outer nucleon with the nuclear core, $\mu$ is the magnetic moment
of the outer nucleon, $r_0=1.2$ fm. The so-called ``best values''
for the parameters of P odd nuclear forces \cite{ddh} result in
$g_p=4.5$ for an outer proton \cite{fks,dfst,fts}. The
$A$-dependence of this constant has been already anticipated from
the picture of a toroidal winding.

Various theoretical predictions (with $g_p=4.5$) for the AMs of
nuclei of those atoms where the PNC effects have been studied
experimentally, are presented in Table 1 (below in this subsection
we follow \cite{dkh}).
\begin{table}
[t]
\begin{center}
\begin{tabular}{|c|c|c|c|c|c|} \hline

&&&&&\\
   & \cite{fks}&\cite{dkt} &\cite{ab}&\cite{dt}
&\cite{hm}\\ &&&&&\\ \hline &&&&&\\ $^{133}$Cs & 0.37
&0.22&---&0.15&0.21\\ &&&&&\\ \hline &&&&&\\ $^{203,205}$Tl &0.49&
0.37&0.24&0.24&0.10\\ &&&&&\\ \hline &&&&&\\ $^{209}$Bi & 0.51 &
 0.30&---&0.15&---\\ &&&&&\\ \hline
\end{tabular}
\end{center}
\begin{center}
Table 1
\end{center}
\end{table}
The analytical estimate (\ref{an}) produces smooth $A^{2/3}$
behaviour (see the first column of Table~1), but certainly
exaggerates the effect due to the assumption that the P odd
contact interaction with the nuclear core extends throughout the
whole localization region of the unpaired nucleon. Indeed, more
refined calculations, already in the single-particle approximation
(SPA) (in the second column of Table 1 we present the results
obtained with the Woods-Saxon potential, including contributions
of contact and spin-orbit currents, which is perhaps the most
advanced calculation in SPA) reveal certain shell effects quite
pronounced in the values of $\kappa$ for Tl and Bi. Both these
nuclei are close to the doubly-magic $^{208}$Pb. However, while
the anapole moment of Tl nucleus in SPA is close to its analytical
estimate, the anapole moment of Bi in SPA differs significantly
both from the analytical formula and from the anapole moment of
Tl. This difference can be attributed to the difference in the
single-particle orbitals for the unpaired proton in Tl and Bi. The
3s$_{1/2}$ wave function in Tl is concentrated essentially inside
the nuclear core, while the 1h$_{9/2}$ wave function in Bi is
pushed strongly outside of it. By this reason the unpaired proton
in Bi ``feels'' in fact much smaller part of the P odd weak
potential. An analogous suppression of the PNC interaction takes
place for the outer 1g$_{7/2}$ proton in Cs.

Various approaches were used as well in the many-body
calculations. In one of them \cite{dt} the random-phase
approximation (RPA) was employed to calculate the effects of the
core polarization. In another approach \cite{hm} large basis
shell-model calculations were performed. However, in the last case
there is a serious problem: the basis necessary to describe
simultaneously the effects of both regular nuclear forces and P
odd ones, is in fact too large. Therefore, some additional
approximations were made in \cite{hm} in order to reduce the size
of the basis space.

Fortunately, the Tl nucleus is a rather special case in the
many-body approach as well. Not only is it close to the
doubly-magic $^{208}$Pb, but its unpaired proton is 3s$_{1/2}$,
but not 1h$_{9/2}$ as in Bi. This makes the effects of the core
polarization here relatively small. Thus the density of states in
Tl is reduced, and an effective Hamiltonian suitable for
shell-model calculations can be constructed \cite{rblp}. This
Hamiltonian was used in \cite{ab} to calculate the anapole moment
of Tl nucleus. The result of \cite{ab} and the RPA result of
\cite{dt} for the thallium coincide, in spite of completely
different descriptions of nuclear forces used in these works to
calculate the core polarization. These results of \cite{ab,dt}
differ essentially from the value obtained in \cite{hm} under
extra assumptions: the closure approximation and further reduction
of a three-body matrix element to the two-body one. It is also
worth mentioning perhaps that in \cite{ab,dt} and \cite{hm}
different parameterizations of the parity violating nuclear forces
have been used.

Thus we believe that the theoretical predictions for the AMs of
nuclei of the present experimental interest, can be reasonably
summarized now, at ``best values'' of P odd constants, as follows:
\beq\label{pre}
\kappa(^{133}{\rm Cs})=0.15-0.21, \quad \kappa(^{203,205}{\rm
Tl})=0.24, \quad \kappa(^{209}{\rm Bi})=0.15.
\eeq
We believe also that there are good reasons to consider these
predictions as sufficiently reliable, at the accepted values of
the P odd nuclear constants.

The comparison of the value (\ref{pre}) for the cesium AM with the
experimental result (\ref{exp}) indicates that the ``best values''
of \cite{ddh} somewhat underestimate the magnitude of P odd
nuclear forces. In no way is this conclusion trivial. The point is
that the magnitude of parity-nonconserving effects found in some
nuclear experiments is much smaller than that following from the
``best values'' (see review \cite{ah}). In all these experiments,
however, either the experimental accuracy is not high enough, or
the theoretical interpretation is not sufficiently convincing. The
experiment \cite{woo} looks much more reliable in both respects.
Therefore, in line with its general physics interest, the
investigation of nuclear AMs in atomic experiments is first-rate,
almost table-top nuclear physics.

It is appropriate perhaps to point out here a problem we still
have not mentioned. The experimental result for the thallium AM,
$\kappa =-0.22\pm 0.30$ \cite{vet}, does not comply with the
theoretical prediction for it presented in (\ref{pre}) (the
disagreement will be even more serious if one assumes that the
nuclear P odd constants are larger than the ``best values'' of
\cite{ddh} as indicated by the measurement of the cesium AM).
Obviously, it is highly desirable for this problem to be cleared
up.

\subsection{Prospects}
The experiments aimed at the detection and measurement of nuclear
AMs are extremely difficult. Therefore, it would be quite
important to find a situation where the AM effects are strongly
enhanced. Unfortunately, up to now nobody could point out any
example of a pronounced enhancement of a nuclear AM by itself.

In principle, the AM can be enhanced in the case when anomalously
close to the ground state of a nucleus there is an opposite-parity
level of the same angular momentum. In this connection, attention
was attracted in \cite{hu, hus} to exotic halo nuclei. In
particular, the exotic neutron-rich halo nucleus $^{11}$Be was
considered therein. In this nucleus the outer odd neutron is in
the state $2s_{1/2}$, its only bound excited level being
$1p_{1/2}$ (the well-known ``inversion of levels''). The
anomalously small energy separation between these two levels of
opposite parity,
\beq\label{int}
|\De E| = E(1p_{1/2}) - E(2s_{1/2}) = 0.32\, {\rm MeV},
\eeq
enhances by itself their P-odd mixing and thus the AM of this
nucleus. As pointed out in~\cite{hu,hus}, the small binding energy
of the odd neutron,
\beq\label{be}
|\De E_0| = 0.50\, {\rm MeV},
\eeq
affects the AM additionally, but in two opposite directions. On
one hand, it suppresses the overlap of the odd-neutron wave
function with the core, and thus suppresses the mixing of the
$2s_{1/2}$ and $1p_{1/2}$ levels due to the weak interaction
operator which looks as
\beq\label{w}
W=\,\frac{G}{\sqrt{2}}\,\frac{g_n}{2m}\,\{\si \p, \rho(r)\}\,;
\eeq
here $g_n$ is the effective constant of the P-odd interaction of
the outer neutron with the nuclear core, $\si$ and $\p$ are the spin and
momentum operators of the outer neutron, and $\rho(r)$ is
the spherically symmetric core density. On the other hand, the
small binding energy enhances the matrix element of $\br$ in the
anapole operator of the neutron
\beq\label{a}
\mathbf{a}=\,\frac{\pi e \mu_n}{m}\,\br \times \si\,;
\eeq
here $\mu_n=-1.91$ is the neutron magnetic moment.

The detailed calculation which takes into account the P-odd mixing
of the ground state with the $1p_{1/2}$ level only, results in the
following value for the effective anapole constant~\cite{hus}:
\begin{equation}\label{b1}
\kappa_1(^{11}{\rm Be})= 0.17 g_n.
\end{equation}
Indeed, this value is 15 times larger than that given by the
estimate (\ref{an}) for $A=11$ (the neutron constant $g_n$ is
poorly known by itself, most probably $g_n\, \lsim \,1$).
Certainly, this enhancement of an AM in a light nucleus would be
of a serious interest, even if its possible experimental
implications are set aside.

However, such a strong enhancement of AM, as given in (\ref{b1}),
in a loosely bound nucleus does not look natural. In particular,
nothing of the kind happens in the deuteron. Even in the limit of
vanishing binding energy, when the energy interval between the
deuteron $s\,$ state and the continuum $p\,$ states tends to zero,
the deuteron AM is not enhanced essentially (see below). As to the
problem of $^{11}$Be discussed here, a strong cancellation between
the contribution of the bound $1p_{1/2}$ state (accounted for in
(\ref{b1})) and that of the continuum (omitted therein) takes
place \cite{adk}. This cancellation results in a serious
suppression of the naive estimate as compared to (\ref{b1}):
\beq
\kappa(^{11}{\rm Be}) \simeq 0.07 g_n.
\eeq

Let us come back now to the above mentioned case of the deuteron
AM. It can be calculated in a closed form in the chiral limit,
i.~e. for the vanishing pion mass, $m_{\pi} \to 0$ \cite{kk} (see
also \cite{sav}). In this limit one can confine for the loosely
bound deuteron to the weak one-pion exchange. We use the
Lagrangians of the strong $\pi$NN interaction and of the weak
P-odd one, $L_s$ and $L_w$, in the form
\begin{equation}\label{s}
L_s\,=\,g\,[\,\sqrt{2}\,(\overline{p}i\gamma_{5}n \,\pi^+
+\overline{n}i\gamma_{5}p\, \pi^-)\,+(\,\overline{p}i\gamma_{5}p
-\overline{n}i\gamma_{5}n)\,\pi^0];
\end{equation}
\begin{equation}\label{w1}
L_w\,=\,\bar{g}\,\sqrt{2}\,i\,(\,\overline{p}n \,\pi^+
-\overline{n}p \,\pi^-).
\end{equation}
Then in the zero-range approximation for the deuteron wave
function, one obtains the following result for its AM
\beq\label{fa}
\va_d^{(0)}\,=\,-\,{e g\bar{g} \over 6 m m_{\pi}} \,{1+\xi \over
(1+2\xi)^2}\,\left(\mu_p-\mu_n- \,{1 \over 3}\right)\bI\,.
\eeq
Here $\mu_p=2.79$ and $\mu_n=-1.91$ are the proton and neutron
magnetic moments, respectively; $\bI$ is the deuteron spin. With
the deuteron binding energy $\ep =2.23$ MeV, the numerical value
of the parameter $\xi=\sqrt{\ep m}/m_{\pi}$ in this formula is
$\xi = 0.32$. As it was mentioned above, even for the vanishing
binding energy $\ep \to 0$ no essential enhancement occurs for the
deuteron AM.

Let us note that in the same chiral limit the nucleon AM can be
calculated in a closed form. It was done in 1980 by I.B.
Khriplovich and A.I. Vainshtein. The result is the same for the
proton and neutron:
\beq\label{nam}
\va_p\,=\,\va_n\,=\,-\,{e g \bar{g} \over 12 m m_{\pi}}\,
\left(1-\,{6 \over \pi}\,{m_{\pi} \over m}\ln{m \over
m_{\pi}}\right) \,\mbox{\boldmath $\sigma$}.
\eeq
To obtain the total deuteron AM, one should combine the
contributions of the nucleon anapoles (\ref{nam}) with (\ref{fa}).

To summarize, no example of a considerable enhancement of nuclear
AMs by internal nuclear effects has been found up to now.

There is however a situation when the manifestation of a nuclear
AM is strongly enhanced by atomic effects. We mean a proposal
  \begin{figure}[t]
\begin{center}
\includegraphics[width=9cm]{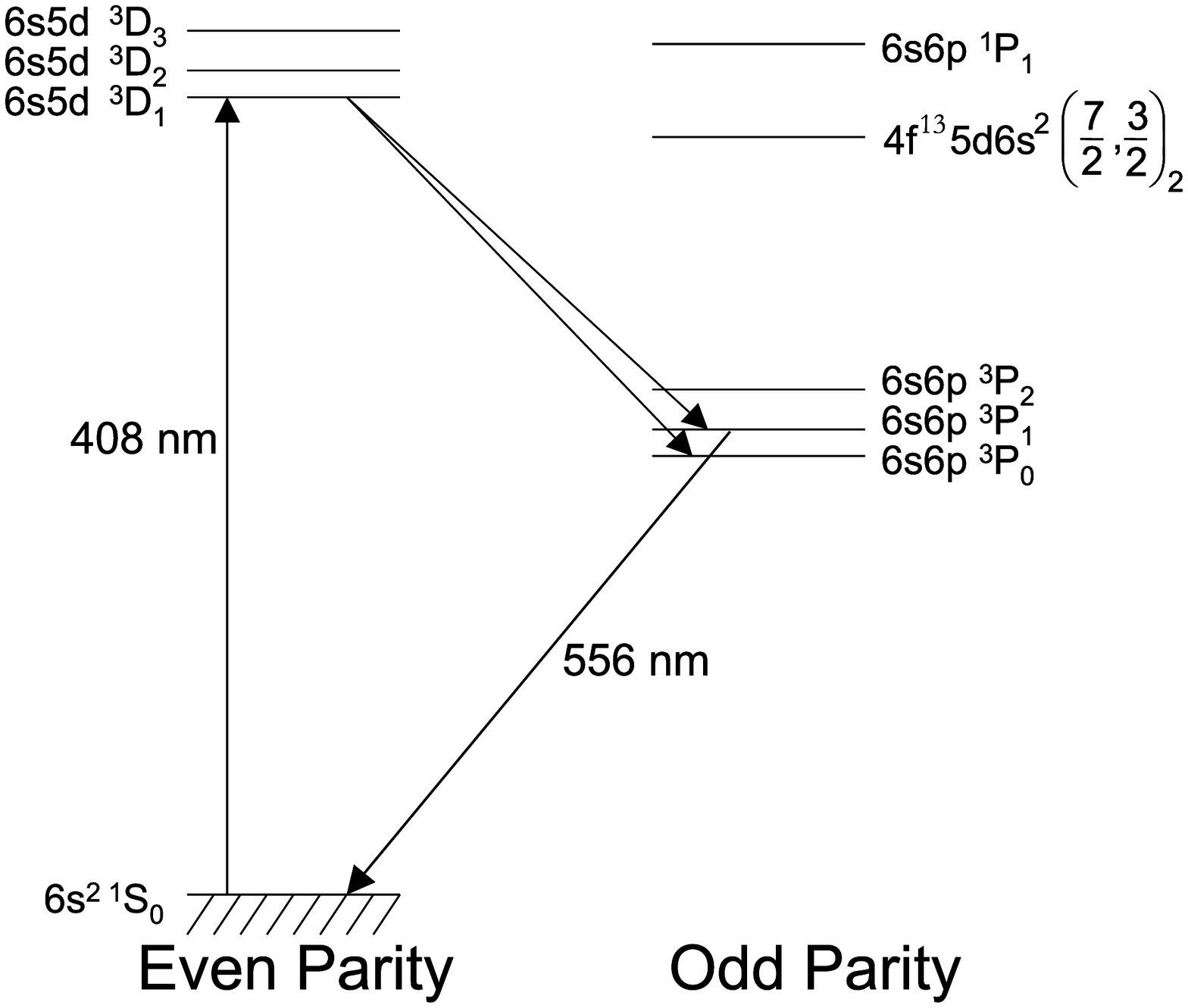}\\
\vspace{0.5cm}
Figure 1
\end{center}
\end{figure}
(which looks at the moment the most promising one) to measure PNC
in the strongly forbidden $6s^2$ $^1S_0 \to 6s5d$ $^3D_1$
transition in ytterbium \cite{dem}. The advantage of this
transition is that the PNC effect in it is more than 100 times
larger than in cesium. The enhancement is due to the fact that the
$^3D_1$ state is close ($\Delta E \simeq 600$ cm$^{-1}$) to a
level of opposite parity $^1P_1$ (see Figure 1) whose composition is $6s6p$ with
a strong admixture of $5d6p$. Due to this $5d6p$ component, there
is large PNC mixing between $^3D_1$ and $^1P_1$. The relatively
simple atomic spectrum of Yb allows one to perform atomic
calculations of the PNC effect with an accuracy about 20\%
[24--26].

Ytterbium has seven stable isotopes between A=168 and A=176. Two
of them, $^{171}$Yb, I=1/2 and $^{173}$Yb, I=5/2, with non-zero
nuclear spin, can be used to measure the AMs. With a valence
neutron in these nuclei, such measurements will be a valuable
complement to the cesium anapole result in the determination of
the PNC nuclear constants.

Moreover, one more transition in Yb, $6s^2$ $^1S_0 \to 6s5d$
$^3D_2$, is of a special interest for the anapole measurements.
PNC mixing between $^3D_2$ and $^1P_1$ states (their separation is
$\Delta E \simeq 300$ cm$^{-1}$) is only due to that P odd
interaction of electrons with nuclear spin which possess $\Delta J
=1$. Thus, in this transition the anapole interaction will be the
main source of P odd effects, rather than a small correction to
the dominant nuclear-spin-independent interaction as it is the
case with $^1S_0 \to ^3D_1$.

Some preliminary spectroscopic measurements in ytterbium, related
to the discussed experiments, have been done. They resulted in the
lifetime of the $^3D_1$ state, 380(30)~ns~\cite{bo}, as well as in
the values of the E2 amplitude of the $^1S_0 \to ^3D_2$
transition, 0.65(3)~$e a_0^2$~\cite{bow}, and of the strongly
forbidden M1 amplitude of the $^1S_0 \to ^3D_1$ transition, $
1.33(26) \times 10^{-4}\, \mu_{{\rm B}}$~\cite{sta}.

\section{Nuclear electric dipole moments}
We start here with the present experimental information on the
EDMs.

\subsection{Elementary particles}

The experimental upper limit on the neutron EDM is
\cite{kfsmi,alt,har}
\beq\label{dn}
d_n < (6-10) \times 10^{-26}\;e\,{\rm cm}.
\eeq
The sensitivity of these experiments can be, hopefully, improved
by 2~--~3 orders of magnitude.

The best result for the electron EDM
\beq\label{de}
d_e = (0.69 \pm 0.74)\times 10^{-27}\;e\,{\rm cm}
\eeq
was obtained in atomic experiment with Tl~\cite{reg}. Hopefully,
this limit can be pushed well into the $10^{-28}\;\;e\;$cm range.

I would like to quote here one more upper limit, that on the muon
EDM~\cite{bai}:
\beq
d_\mu < 10^{-18}\;e\,{\rm cm}.
\eeq
An experiment was recently proposed to search for the muon EDM
with the sensitivity of $10^{-24}\;e\,$cm~\cite{yann}. We will
come back to this proposal later.

The predictions of the Standard Model are, respectively:
\beq
d_n \sim 10^{-32} - 10^{-31} \;e\,{\rm cm};
\eeq
\beq
d_e < 10^{-40}\;e\,{\rm cm};
\eeq
\beq
d_\mu < 10^{-38}\;e\,{\rm cm}.
\eeq

\subsection{Atoms and nuclei}

The best upper limit on EDM of anything was obtained in atomic
experiment with $^{199}$Hg~\cite{rom}. The result for the dipole
moment of this atom is
\beq\label{hg}
d(^{199}{\rm Hg}) < 2.1 \times 10^{-28}\;e\,{\rm cm}.
\eeq
Unfortunately, the implications of the result (\ref{hg}) are
somewhat less impressive, due to the electrostatic screening of
the nuclear EDM in this essentially Coulomb system. The point is
that in a stationary state of such a system, the total electric
field acting on each particle must vanish. Thus, an internal
rearrangement of the system's constituents gives rise to an
internal field $\E_{int}$ that exactly cancels $\E_{ext}$ at each
charged particle; the external field is effectively switched off,
and an EDM feels nothing \cite{pura,gale,schi}.

As to heavy paramagnetic atoms, due to magnetic interactions
increasing rapidly with $Z$, the electron EDM therein is not
suppressed, but enhanced as $Z^3\al^2$ \cite{san}. In particular,
in Tl where the limit (\ref{de}) was obtained, this enhancement
factor reaches $\;$-- 585 \cite{ke}.

However, in the present case of mercury, observable T odd effects
are due to the finite size of the nucleus which is small on the
atomic scale. To explain how these effects arise here, let us
expand the nuclear charge density in powers of the proton
coordinates:
\[
\rho({\bf r}) =e \sum_p  \delta({\bf r} - {\bf r}_p)=e \sum_p
\delta({\bf r}) - e \sum_p  r_p^i \nabla_i \delta({\bf r})
\]
\beq \label{chd}
+ \frac{1}{2}e \sum_p  r_p^i r_p^j \nabla_i  \nabla_j \delta({\bf
r}) - \frac{1}{6} e \sum_p  r_p^i r_p^j r_p^k \nabla_i  \nabla_j
\nabla_k   \delta({\bf r}) + \ldots
\eeq
Going over now to the operators of quadrupole and octupole
moments,
\[
Q_{ij}= e \sum_p  r_p^i r_p^j -\frac{1}{3}\delta_{ij}e \sum_p
r_p^2\,,
\]
\[O_{ijk} = e \sum_p \left[ r_p^i r_p^j r_p^k-
\frac{1}{5}(r_p^i\delta_{jk}+ r_p^j\delta_{ik}+ r_p^k
\delta_{ij})r_p^2\right],
\]
we rewrite formula (\ref{chd}) as
\[
 \rho({\bf r}) = Ze \delta({\bf r}) + \frac{1}{6} Ze \langle r^2\rangle \Delta \delta({\bf r})  - \frac{1}{10}e \sum_p
r_p^i{\bf r}_p^2 \nabla_i \Delta  \delta({\bf r})
\]
\beq \label{chd1}
- d_i\nabla_i  \delta({\bf r}) + \frac{1}{2} Q_{ij} \nabla_i
\nabla_j \delta({\bf r}) - \frac{1}{6} O_{ijk} \nabla_i
\nabla_j\nabla_k  \delta({\bf r}) + \ldots\, ;
\eeq
in this expression $\langle r^2\rangle$ is the mean squared charge
radius of the nucleus.

The term $ - e \sum_p  r_p^i \nabla_i \delta({\bf r})$ in the
charge density generates P odd, T odd correction to the
electrostatic potential. To take into account the mentioned
electrostatic screening, we have to subtract from this correction
the term
\[
(\bd \cdot \Na) \int \frac{\rho({\bf r})}{|{\bf R}- {\bf r}|} d^3{\bf r},
\]
where $\bd$ is the EDM of the nucleus. In this way we obtain the
following P odd, T odd term in the electrostatic potential:
\beq \label{PTpot}
\phi_{PT}({\bf R}) = - 4\pi ({\bf S}\cdot \Na )\delta({\bf R}).
\eeq
Here ${\bf S}$ is the so-called Schiff moment
\beq \label{Sch}
{\bf S} = \frac{1}{10}\,e \sum_p {\bf r}_p
\left(r_p^2-\frac{5}{3}\langle r^2\rangle\right).
\eeq

The contribution to the Schiff moment from the electric dipole
moments $d_N$ of nucleons can be found as follows. Equation
(\ref{Sch}) is valid for any system of point-like charges. Let us
split the sum in it into the sum over coordinates ${\bf r}_N$ of
the centres of mass of nucleons and the sum over coordinates
$\mbox{\boldmath$ \rho$}_i$ of point-like charged constituents
(say, quarks) of the nucleons:
\begin{equation} \label{Sch1}
{\bf S} = \frac{1}{10} \sum_N \sum_i  e_i \left(({\bf r}_N +
\mbox{\boldmath$ \rho$}_i )^2 - \frac{5}{3}\langle
r^2\rangle\right) ({\bf r}_N + \mbox{\boldmath$ \rho$}_i).
\end{equation}
Combining terms of the zeroth and first order in $
\mbox{\boldmath$ \rho$} $ and taking into account that $\sum_i e_i
= e_N$, $\sum_i e_i \mbox{\boldmath$ \rho$}_i = {\bf d}_N$, we
present the arising expression for the Schiff moment as a sum of
two contributions. The first one is similar to (\ref{Sch}), and
originates from the P odd, T odd nucleon-nucleon interaction:
\begin{equation} \label{Sch1a}
{\bf S_1} = \frac{1}{10} \sum_N e_N {\bf r}_N \left(r_N^2 -
\frac{5}{3}\langle r^2\rangle \right);
\end{equation}
here  $e_N=|e|$ for the proton and vanishes for the neutron. The
second contribution is due to the internal dipole moments of the
nucleons
\beq     \label{Sch1b}
{\bf S_2} = \frac{1}{6} \sum_N {\bf d_N} \left(r_N^2 - \langle
r^2\rangle \right)+ \frac{1}{5} \sum_N \left( {\bf r}_N ({\bf r}_N
\cdot {\bf d}_N) - {\bf d}_N r_N^2/3 \right).
\end{equation}

If one ascribes the atomic dipole moment to the EDM of the valence
neutron in the even-odd nucleus $^{199}$Hg, the corresponding
upper limit on the neutron EDM is \cite{ds}
\[
d_n < 4 \times 10^{-25}\;e\,{\rm cm}.
\]
It is few times worse than the direct one (\ref{dn}).

On the other hand, though the $^{199}$Hg nucleus in the
simple-minded shell model has an odd neutron above a spherically
symmetric core, the account for many-body effects allows one to
derive an upper limit on the proton EDM \cite{ds}
\beq\label{dp}
d_p < 5.4 \times 10^{-24}\;e\,{\rm cm}.
\eeq
This is about an order of magnitude better than the limit obtained
from the experiment with TlF molecule \cite{tlf}.

It has been demonstrated, however, that the dipole moments of
nuclei induced by the T and P odd nuclear forces can be about two
orders of magnitude larger than the dipole moment of an individual
nucleon~\cite{sfk}. In the simplest approximation of the shell
model, where the nuclear spin coincides with the total angular
momentum of an odd valence nucleon, while the other nucleons form
a spherically symmetric core with the zero angular momentum, the
effective T and P odd single-particle potential for the outer
nucleon is
\beq\label{W}
W = \,\frac{G}{\sqrt{2}}\,\frac{\xi}{2m_p}\,\si\cdot\Na \rho(r).
\eeq
Here $\xi$ is a dimensionless constant characterizing the strength
of the interaction in units of the Fermi weak interaction constant
$G$; $m_p$, $\si$, and $\br$ are the mass, spin and coordinate of
the valence nucleon, respectively.

A simple closed form for the nuclear EDM induced by interaction
(\ref{W}) can be derived as follows. Since the profiles of the
nuclear core density $\rho(r)$ and the potential $U(r)$ for the
outer nucleon are similar, let us assume that they coincide
exactly:
\[
\rho(r)=U(r)\,\frac{\rho_0}{U_0}.
\]
Then the perturbation (\ref{W}) can be rewritten as
\beq
W(\br) = \la\si\cdot\Na U(r), \quad
\la=\,\frac{G}{\sqrt{2}}\,\frac{\xi}{2m_p}\,\frac{\rho_0}{U_0}=-2\times
17^{-21}\,\xi\,{\rm cm}.
\eeq
Accordingly, the total potential in which the nucleon moves is
\beq
\tilde{U}(\br) = U(r)+W(\br) = U(r)+\la\si\cdot\Na U(r) =
U(|\br+\la\si|). \eeq In this potential, it is obvious that the
wave function of the external nucleon becomes
\beq
\tilde{\psi}(\br) = \psi(\br+\la\si) =
(1+\la\si\cdot\Na)\psi(\br),
\eeq
where $\psi(\br)$ is its unperturbed value. It is a simple problem
now to obtain the following closed expression for the nuclear EDM:
\beq
d_N = - e(q-Z/A)\la\,\frac{1/2-K}{I+1}.
\eeq
Here $Z$, $A$, and $I$ are the nuclear charge, atomic number, and
spin, respectively; $K=(l-I)(2I+1)$, where $l$ is the orbital
angular momentum of the outer nucleon; $q=1$ or $0$ for an
external proton or neutron, respectively. It is curious that this
result of a simple-minded analytical approach is very close to
that of the best numerical single-particle calculations.

The characteristic value of the thus induced nuclear EDM is
\beq\label{dN}
d_N \sim 10^{-21}\,\xi \;e\,{\rm cm}.
\eeq
Being interpreted in terms of the CP odd nuclear forces, the
experimental result (\ref{hg}) leads to the following upper limit:
\beq\label{xi}
\xi < 0.5\times 10^{-3}.
\eeq

There are good reasons to assume that the exchange by
$\pi^0$-meson is the most efficient mechanism of generating CP odd
nuclear forces. This is due to the large value of the strong $\pi
NN$ coupling constant $g_s = 13.5$ and to the small pion mass, as
well as to the fact that outer proton and neutron orbitals in
heavy nuclei are quite different. The P odd, T odd effective $\pi
NN$ Lagrangians are conveniently classified by their isotopic
properties \cite{bar,hahe,herc}:
\begin{equation}\label{0}
\De T =0. \quad L_0\,=\,\bar{g}_0\,[\,\sqrt{2}\,(\overline{p}n
\,\pi^+ +\overline{n}p\, \pi^-)\,+(\,\overline{p}p
-\overline{n}n)\,\pi^0];
\end{equation}
\begin{equation}\label{1}
|\De T| = 1. \quad L_1\,=\,\bar{g}_1\,\overline{N}N \,\pi^0 =
\,\bar{g}_1\,(\,\overline{p} p \,+ \overline{n}n )\,\pi^0;
\end{equation}
\begin{equation}\label{2}
|\De T| =2. \quad L_2\,=\,\bar{g}_2\,[\,\sqrt{2}\,(\overline{p}n
\,\pi^+ +\overline{n}p\, \pi^-)\,- 2(\,\overline{p}p
-\overline{n}n)\,\pi^0].
\end{equation}
With the strong interaction Lagrangian given by formula (\ref{s}),
it can be easily seen that in heavy nuclei, at least in the
single-particle approximation, the contributions of the P odd, T
odd interactions $L_0$, $L_2$, which conserve isospin and change
it by 2, are proportional to $N-Z$ and thus are suppressed as
compared to $L_1$ interaction changing isospin by 1, which is
proportional to $N+Z=A$. For the present case of the $^{199}$Hg nucleus there is
an additional occasional cancellation in the isoscalar contribution. In this way
the limit (\ref{xi}) can be transformed to
\beq\label{g1}
g_1 < 0.5 \times 10^{-11}.
\eeq

The Standard Model (SM) prediction for this constant is \cite{kky}
\beq
g_1 \sim 10^{-17}.
\eeq
Thus, the theoretical predictions of the SM for dipole moments and
CP odd nuclear forces are about six orders of magnitude below the
present experimental upper limits on them. In fact, this means
that the searches for electric dipole moments now, at the present
level of accuracy, are extremely sensitive to possible new
physics.

Theoretical models of CP violation are too numerous to discuss all
of them, and most of them have too many degrees of freedom. It is
convenient therefore to proceed in a phenomenological way: to
construct CP odd quark-quark, quark-gluon, and gluon-gluon
operators of low dimension, and find upper limits on the
corresponding coupling constants from the experimental results for
$d_n$ and d($^{199}$Hg). The analysis performed
in~\cite{kky,khkh}, has demonstrated that the limits on the
effective CP odd interaction operators obtained from the neutron
and atomic experiments are quite comparable. These limits are very
impressive. All the constants are several orders of magnitude less
than the usual Fermi weak interaction constant $G$. In particular,
these limits strongly constrain some popular models of CP
violation, such as the model of spontaneous CP violation in the
Higgs sector, and the model of CP violation in the supersymmetric
S0(10) model of grand unification.

\section{Electric dipole moments, storage rings\\ and linear traps}
The various upper limits on EDMs set so far, constitute a valuable
contribution to elementary particle physics and to our knowledge
of how the Nature is arranged; the null results obtained so far
are important. But it is only natural to think of essential
progress in the field, of finding a positive result, of eventually
discovering permanent electric dipole moment. In particular, it
would be tempting to get rid of the electrostatic screening of
nuclear EDMs. So, let me add to the above stories, a new one. It
should be started with the discussion of

\subsection{Idea of new muon EDM experiment}
The idea is to search for the muon EDM in a storage ring, with
muons in it having natural longitudinal polarization~\cite{yann}.
An additional spin precession due to the EDM interaction with
external field should be monitored by counting the decay
electrons, their momenta being correlated with the muon spin, due
to parity nonconservation in the muon decay. The frequency $\vom$
of the spin precession with respect to the particle momentum in
external magnetic and electric fields, $\B$ and $\E$, is
\beq\label{prec}
\vom=-\frac{e}{m}\left[a\B-a\,\frac{\ga}{\ga+1}\bv(\bv\cdot\B)-
\left(a-\frac{1}{\ga^2-1}\right)\bv\times\E\right]
\eeq
\[
-\eta\,\frac{e}{m}\left[\E-\,\frac{\ga}{\ga+1}\bv(\bv\cdot\E)
+\bv\times\B\right].
\]
Here the anomalous magnetic moment $a$ is related to the
$g$-factor as follows: $a = g/2 - 1$ (for muon $a = \al/2\pi)$;
$\bv$ is the particle velocity; $\ga = 1/\sqrt{1-v^2}$. The last
line in this formula describes the precession due to the EDM $d$,
the dimensionless constant $\eta$ being related to $d$ as follows:
\[
d =\,\frac{e}{2m}\,\eta.
\]
This last line can be obtained from the terms proportional to the
anomalous magnetic moment in (\ref{prec}) by substituting $a \to
\eta$ and changing to dual fields: $\B \to \E, \;\;\; \E \to -
\B$. Expression (\ref{prec}) simplifies in the obvious way for
$(\bv\cdot\B) = (\bv\cdot\E) = 0$. Just this case is considered
below.

The idea of~\cite{yann} is to compensate for the usual spin
precession in the vertical magnetic field $\B$ by the precession
in a radial electric field $\E$, i. e. to choose $\E$ in such a
way that the first line in (\ref{prec}) vanishes at all. Then the
spin precession with respect to momentum is due only to the EDM
interaction with the vertical magnetic field, and since electric
fields in a storage ring are much smaller than magnetic ones, it
reduces to
\beq
\vom = \vom_e = -\,\frac{e}{m}\,\eta\, \bv \times \B.
\eeq
In a nutshell, due to the EDM interaction, the muon spin precesses
around the motional electric field $\bv \times \B$. In this way
the spin acquires a vertical component which linearly grows with
time. The P odd correlation of the decay electron momentum with
the muon spin leads to the difference between the number of
electrons registered above and below the orbit plane.

In~\cite{yann}, it is stated that the sensitivity to the muon EDM
can be improved in the planned experiment by six orders of
magnitude, to the level of $10^{-24}\;e\,$cm.

But after all, the useful signal here is due to the spin
precession in the electric field in the muon rest frame. So, the
question is: what about the screening of the electric field which
is responsible in particular for reducing the record-breaking
result (\ref{hg}) to much more modest bound on $d_n$? The
explanation is as follows. All previous EDM experiments consisted
essentially in the measurements of frequencies, i. e. referred to
stationary states. In the present case the spin precession itself
is being measured, i. e. this is completely nonstationary
situation. Thus there is no screening suppression here.

\subsection{Nuclear electric dipole moments at ion storage rings}
In the same way one can search for an EDM of a polarized
$\be$-active nucleus in a storage ring~\cite{khr}. In this case as
well, the precession of nuclear spin due to the EDM interaction
can be monitored by the direction of the $\be$-electron momentum.
$\be$-active nuclei have serious advantages as compared to muon.
The life time of a $\be$-active nucleus can exceed by many orders
of magnitude that of a muon. The characteristic depolarization
time of the ion beam can reach few seconds, and is also much
larger than the muon life time, which is about $10^{-6}$~s.
Correspondingly, the angle of the rotation of nuclear spin, which
is due to the EDM interaction and which accumulates with time, may
be also by orders of magnitude larger than that of a muon. By the
same reason of the larger life time, the quality of an ion beam
can be made much better than that of a muon beam.

However, necessary conditions here are also quite serious.

First of all, to make realistic the mentioned compensation of the
EDM-independent spin precession by a relatively small electric
field, the effective nuclear $g$-factor should be close to 2 (as
this is the case for the muon). For a nucleus with the total
charge $Ze$, mass $Am_p$, spin $I$, and magnetic moment $\mu$, the
effective anomalous magnetic moment is
\[
a=\,\frac{g}{2}\,-1=\,\frac{A}{Z}\,\frac{\mu}{2I}\,-1.
\]
Some fine-tuning of $a$ is possible sometimes  by taking, instead
of a bare nucleus, an ion with closed electron shells. An accurate
formula for the anomaly of an ion with the total charge $z$, is
\beq
a=\,\frac{A}{2z}\,\frac{\mu}{I}\,-1.00722+\,\frac{\De}{Am_p}\,
-\,\frac{z}{A}\,\frac{m_e}{m_p}.
\eeq
It includes the correction for the atomic mass excess $\De$.

It is more practical perhaps to confine, in line with bare nuclei,
to helium-like ions, i.~e. to $z=Z,\; Z-2$. One of the reasons is
the same electrostatic screening. It is complete for a neutral
atom, and for an ion it is only partial, being proportional to the
number of electrons $Z-z$. So, the smaller this number is, the
better for our problem.

Some ions which look at the moment promising from the point of
view of the EDM searches are presented in the Table. More complete
list, comprising about 30 candidates, can be found in~\cite{khr}.

\begin{table}
\begin{center}
\begin{tabular}{cccccrcc}
\hline
 &$I^{\pi}\to I^{\pi'}$&$\mu$&$z$&$a\times10^3$&$t_{1/2}$&$Q$(barn)& branching\\
\hline $^{131}_{\;\,53}$I & $7/2^+\to 5/2^+$ &2.742(1)&51&
$-$1.9(0.4)&8.0 d&$-$0.40&90\%\\ $^{133}_{\;\,53}$I & $7/2^+\to
5/2^+$ &2.856(5)&53& 16(2)&21 h&$-$0.27&83\%\\ $^{139}_{\;\,55}$Cs
& $7/2^+\to 7/2^-$ &2.696(4)&53& 2(1)&9.3 m&$-$0.075&82\%\\
$^{223}_{\;\,87}$Fr & $3/2(^-)\to 3/2^-$ &1.17(2)&87&
$-$7$\pm$20&22 m&1.2&67\%\\ \hline
\end{tabular}
\end{center}
\begin{center}
Table 2
\end{center}
\end{table}
The errors in the values of anomalous magnetic moments $a$,
presented in the Table, correspond to the experimental errors in
values of $\mu$. The $\be^-$ branchings are indicated in the last
column.

Let us note that a background due to nuclear quadrupole moments,
even as large as $Q \sim 1$ barn, is not dangerous at reasonable
parameters of a storage ring even for the EDM sensitivity as high
as $10^{-26}\;e\,$cm.

It looks at the moment that the most serious problem for both muon
and nuclear EDM experiments consists in extremely strict demands
on the alignment of the radial electric field. According to
I.~Koop and Y.~Semertzidis, its spurious vertical component is
bounded as follows:
\beq\label{al}
\frac{E_z}{E_r}\, < \eta\,\frac{v}{a}\,.
\eeq
For the muon experiment this ratio should not exceed $10^{-8}$.
Such an accuracy is difficult to attain. The situation is better
for ions at the same ratio $v/a$. The point is that with the same
EDM value, $\eta$ for a nucleus is larger by a factor
$(A/z)(m_p/m_\mu)\sim 20$. However, even for ions the problem is
very serious.

An excellent idea by I. Koop is to work here without electric
field at all. As to muons, their life time is short, $2\times
10^{-6}$ s. So, during the measurement their spins anyway will
make only few turns due to $a$. Therefore, in this case it looks
reasonable to make the measurement time smaller, i. e. to
sacrifice an order of magnitude in sensitivity, but to get rid of
the formidable problem. As to ions, the idea works differently.
Here the residual spin precession due to $a$ can be used for
modulation of the useful signal, i. e. of the spin precession due
to the EDM. The modulation of the signal partly compensates for
the loss in its magnitude.

Now, if a sufficiently large EDM signal can be attained, i. e. if
the angle of the spin rotation can reach, say, a milliradian, one
could think about an experiment with stable nuclei or nuclei of a
large life time. Their polarization could be measured in
scattering experiments (the idea advocated by Y. Semertzidis and
A. Skrinsky). Here the deuteron is a prominent candidate
\cite{khr1} in spite of its not so small anomaly $a=-\,0.143$. The
deuteron fluxes are attainable exceeding $10^{12}$ particles per
second, with high degree of polarization. The deuteron
polarization is measured now with an accuracy $\sim 10^{-2}$. From
the theoretical point of view, the calculation of its EDM $d_d$ is
a relatively clean problem. If induced by P odd, T odd nuclear
forces, $d_d$ can be calculated in the same way as the deuteron AM
with the following result \cite{kk}:
\beq\label{dd}
{\mathbf d}_d\,=\,-\,{e g\bar{g}_1 \over 12 \pi m_{\pi}} \,{1+\xi
\over (1+2\xi)^2}\,\bI\,.
\eeq
Numerically, it is
\beq\label{ded}
d_d= - 2.4\times 10^{-14}\,g_1\;e\,{\rm cm}.
\eeq
Calculations with more realistic (than zero-range approximation)
deuteron wave functions allows one to estimate the accuracy of
this result as 30\%. Thus, the deuteron EDM is due essentially to
the same P odd, T odd $\pi NN$ constant $g_1$ as the EDM of a
heavy nucleus.

On the other hand, the deuteron EDM can arise due to the proton
and neutron dipole moments. With the deuteron being essentially a
$^3S_1$ bound state, this contribution to its EDM is
\beq\label{dpn}
d_d(n,p)= d_p + d_n.
\eeq

If one assumes that the nucleon dipole moments are also due to the
P odd, T odd $\pi NN$ interaction, they can be calculated in the
chiral limit as well \cite{cr}. The correspondig results, written
in terms of the constants introduced in formulae (\ref{s}),
(\ref{0}), (\ref{2}), are
\beq\label{pn}
d_n = - d_p = \,\frac{e}{m}\,\frac{g
(\bar{g}_0+\bar{g}_2)}{4\pi^2}\,\ln\frac{m}{m_{\pi}}.
\eeq

The idea to use a dedicated deuteron storage ring for measuring
$d_d$ with the sensitivity of $10^{-26}\;e\,{\rm cm}$ is now
seriously discussed by experimentalists (see
http://www2.bnl.gov/~muonedm/). This experiment would be a big
leap forward in the investigation of the nature of CP violation.

On the one hand, it would give immediately the information on the
sum of the EDMs $d_p+d_n$ on the same level of $10^{-26}\;e\,{\rm
cm}$, a gain by an order of magnitude as compared to (\ref{dn}),
(\ref{dp}).

On the other hand, being interpreted in terms of T and P odd
nuclear forces, it would correspond in virtue of (\ref{ded}) to
the sensitivity for $g_1$ an order of magnitude better than the
result (\ref{g1}) of the mercury experiment.

The great importance of the deuteron EDM experiment is obvious.\\

And at last

\subsection{Linear electrostatic trap for heavy polar molecules}
We have mentioned already the formidable problem of the electric
field alignment (see (\ref{al})), perhaps the most serious one for
the storage ring EDM experiments. A cardinal solution of the
problem of spurious spin rotation would be to get rid of external
magnetic fields at all and to make the velocity as small as
possible to suppress the motional magnetic field in the rest
frame. Both these aims are reached, at least in principle, in the
proposal to use linear electrostatic trap for cold polar molecules
\cite{koop}.

The trap looks as a long capacitor where cold molecules move along
$z$ axis. The electric field $\E$ is constant on the $z$ axis and
directed along $y$, it changes only at the ends, thus forming a
longitudinal potential well. On the other hand, the profile of
$\E$ changes in the $xy$ plane in such a way as to guarantee a
sort of strong focusing for molecules aligned along $y$ and moving
along $z$ axis. Due to the essentially one-dimensional motion of
the molecules, one also gets rid (at least, to first
approximation) of one more serious problem, that of the inertial
dragging of spin.

The nuclear spin (or electron one in a radical) precesses in the
$xz$ plane due to the EDM interaction with the effective electric
field. It is not the capacitor field $E_y$, but the intermolecular
one which is huge, it can amount to $10^9$ V/cm. This last
advantage of experiments with polar molecules was in fact pointed
out long ago \cite{sa}.

Obviously, the potential advantages of this idea are huge, and it
well deserves serious experimental investigations.

\end{document}